\documentclass[fleqn,usenatbib]{mnras}

\usepackage{newtxtext,newtxmath}

\usepackage[T1]{fontenc}

\DeclareRobustCommand{\VAN}[3]{#2}
\let\VANthebibliography\thebibliography
\def\thebibliography{\DeclareRobustCommand{\VAN}[3]{##3}\VANthebibliography}

\usepackage{graphicx}	
\usepackage{amsmath}	

\newcommand{\diso}{$d_{\rm iso}$}
\newcommand{\lcdm}{$\Lambda$CDM}
\newcommand{\msol}{$\mathrm{M}_{\odot}$}

\title[Self-regulated SMBH seeding from Pop III.1-s]{The formation of supermassive black holes from Population III.1 seeds. IV.\\Self-regulated seeding from supermassive star ionizing feedback}

\author[M. A. Petkova et al.]{
Maya~A.~Petkova,$^{1}$\thanks{E-mail: maya.petkova@chalmers.se (MAP)}
Jonathan~C.~Tan,$^{1,2}$
Jasbir~Singh,$^{3}$
Vieri~Cammelli,$^{4}$
Mahsa~Sanati,$^{5}$
\newauthor
Benjamin~Keller,$^{6}$
Pierluigi~Monaco,$^{7,8,9,10}$
and Devesh~Nandal$^{11}$
\\
$^{1}$Dept. of Physics and Astronomy, Chalmers University of Technology, SE-412 96 Gothenburg, Sweden\\
$^{2}$Dept. of Astronomy, Univ. of Virginia, Charlottesville, VA, USA\\
$^{3}$INAF - Osservatorio Astronomico di Brera, via Brera 20, I-20121 Milano, Italy\\
$^{4}$Department of Physics, Informatics \& Mathematics, University of Modena \& Reggio Emilia, via G. Campi 213/A, 41125, Modena,
Italy\\
$^{5}$Department of Physics, University of Oxford, Keble Road, Oxford OX1 3RH, UK\\
$^{6}$Department of Physics and Materials Science, University of Memphis, Memphis, TN 38152, USA\\
$^{7}$Dipartimento di Fisica, Sezione di Astronomia, Universit\`{a} degli Studi di Trieste, via G.B. Tiepolo 11, I-34131, Trieste, Italy\\
$^{8}$INAF - Osservatorio Astronomico di Trieste, via G.B. Tiepolo 11, I-34131, Trieste, Italy\\
$^{9}$INFN, Sezione di Trieste, Via Valerio 2, 34127 Trieste TS, Italy\\
$^{10}$IFPU, Institute for Fundamental Physics of the Universe, via Beirut 2, 34151 Trieste, Italy\\
$^{11}$Center for Astrophysics, Harvard and Smithsonian, 60 Garden St, Cambridge, MA 02138, USA\\
}

\date{Accepted XXX. Received YYY; in original form ZZZ}

\pubyear{\the\year{}}

\begin{document}
\label{firstpage}
\pagerange{\pageref{firstpage}--\pageref{lastpage}}
\maketitle

\begin{abstract}
Supermassive Population III.1 stars, i.e., formed from pristine, metal-free gas leading to conditions where dark matter annihilation heating is significant, have been proposed as the progenitors of supermassive black holes (SMBHs) in the early universe ($z \sim 20-40$). Since such Pop III.1 stars only form from non-irradiated gas in dark matter minihalos, they are predicted to appear isolated from each other and other sources of feedback. The previous papers in this series used the isolation distance of Pop III.1 stars as a free parameter to seed SMBH in cosmological simulations of dark matter halos.
Here we develop a feedback-regulated model of Pop III.1 isolation, based on the growth of HII regions around each Pop III.1 star and lower-mass, irradiated Pop III.2 stars. 
Our model considers the time delay between the formation of a minihalo and its Pop III.1 star, R-type expansion of HII regions that expand into the intergalactic medium (IGM), and the redshift dependence of Str\"omgren spheres for longer-lived ionizing sources. For a fiducial Pop III.1 star H-ionizing photon luminosity of $10^{53}\:{\rm s}^{-1}$ and lifetime of $10\:$Myr we find an R-type HII region radius of $R_{\rm R}\simeq1.3\:$cMpc, approximately independent of redshift. The median formation redshift is $\sim20$, with the process essentially complete by $z\sim16$. The overall number density of SMBHs produced in this model is then $n_{\rm SMBH}\simeq 3 \phi_V/(4\pi R_{\rm R}^3)\simeq 0.2\:{\rm cMpc}^{-3}$. We also discuss predictions for the abundance of binary SMBHs, which may appear as dual active galactic nuclei (AGN; $\lesssim 0.3\%$ for $z>6$), and SMBH binary merger rates, measurable by the forthcoming LISA mission.
\end{abstract}




\section{Introduction}\label{sec:intro}

Supermassive black holes (SMBHs) are found in the nuclei of most large and some small galaxies, and are thought to have a major impact on galaxy evolution via their accretion-powered feedback. However, the origin of SMBHs remains poorly understood \citep[e.g.,][]{Volonteri2021}. 
SMBHs are characterized by having masses $\gtrsim 10^5\:M_\odot$, much greater than the mass scale of black holes produced from normal galactic stellar populations, i.e., $\sim 3-100\:M_\odot$. The discovery of SMBH-powered active galactic nuclei (AGN) at high redshift ($z\sim 7-10$; e.g. \citealt{Mortlock2011,Yang2020,Wang2021,Bogdan2024,Maiolino2024a}) greatly constrains the amount of time available for accumulating these huge masses. If the initial seeds of the observed SMBHs are ``light'' (i.e., $\sim100\:M_\odot$), they would require \textit{sustained} super-Eddington accretion, the feasibility of which is currently not supported by either observations \citep[e.g.,][]{Eilers2020,Yang2021,Harikane2023,Maiolino2024b,Zhang2026} or numerical simulations \citep[e.g.,][]{Jeon2023,Sanati2025b}. In addition, the current limited evidence for intermediate mass black holes \citep[IMBHs; e.g.,][]{Greene2020,Mummery2025} points in favour of SMBH formation models that begin with ``heavy'' seeds, i.e., initial masses $\gtrsim10^4\:M_\odot$.

Various physical models have been put forward to explain the origins of SMBHs \citep[e.g.,][]{Rees1978,Volonteri2021}. One such model that may produce heavy seeds is ``direct collapse'', in which the baryonic matter of a metal-free, irradiated, atomically-cooled halo, i.e., with $\gtrsim10^7\:M_\odot$, collapses rapidly to form a single SMBH \citep[e.g.,][]{Bromm2003,Chon2016,Wise2019,OBrennan2025}. This scenario requires that the central protostar continues to accrete significant mass within a time that is shorter than its current Kelvin-Helmholz contraction timescale, so that its surface remains large and relatively cool. To avoid fragmentation, the halo needs to be externally irradiated by a large enough FUV flux to prevent the formation of molecular hydrogen. Due to these special conditions, the favourable formation sites are within satellite galaxies, and the predicted number density of SMBHs formed in this way is too low ($\lesssim 10^{-4}{\rm cMpc}^{-3}$) \citep{Chon2016,Wise2019,OBrennan2025} 
to explain the overall abundance of SMBHs seen at $z=0-7$ of $n_{\rm SMBH}>10^{-2}\:{\rm cMpc}^{-3}$ \citep[][]{Hayes2024,Cammelli2025b}.

Another SMBH formation mechanism that has been considered is via runaway stellar mergers \citep[e.g.,][]{PortegiesZwart2004}. These are proposed to occur in young and dense clusters, where massive (proto-)stars may reach the centre of the cluster before exploding as supernovae, and increase the collision rate drastically \citep[e.g.,][]{Freitag2006,Schleicher2023}. This can result in a seed mass of the order $\sim 200 - 10^3$~\msol{}. However, the occurrence of the necessary conditions for this formation scenario is not well known and hence the predicted SMBH numbers on a cosmological scale are highly uncertain. Furthermore, this mechanism would produce large numbers of IMBHs, which, as mentioned, are not observed.

Finally, Population III stars, i.e., metal-free stars forming from $\sim10^6\:M_\odot$ ``minihalos'', have also been considered as SMBH seeds \citep[e.g.,][]{Madau2001}. While the very low metallicity of the initial star-forming gas is expected to allow these stars to reach higher masses than their present day counterparts, a typical mass range for Pop III stars is estimated to be $\sim 10 - 10^3\:M_\odot$ set by ionizing feedback from the protostar as it contracts towards a main sequence configuration \citep[e.g.,][]{Tan2004,McKee2008,Tan2010,2011Sci...334.1250H,Greif2011,Hirano2014,Jaura2022,Klessen2023}. 

However, it has been proposed that much higher masses could be reached if there is an extra heating source within the protostar coming from Weakly Interacting Massive Particle (WIMP) dark matter self-annihilation \citep{Spolyar2008,Freese2008,Natarajan2009}. This results in stars that have larger radii (i.e., $\gtrsim1\:$AU) and thus relatively low surface temperatures ($\sim10^4~\rm K$) and ionizing luminosities, allowing for mass accretion to be sustained until the supermassive regime is reached \citep[$\sim 10^4-10^5\:M_\odot$;][]{Rindler-Daller2015,Nandal2025,Topalakis2025}. 

In this paper series, we focus on an SMBH seeding model based on ``Pop III.1 stars'' \citep{Banik2019,Singh2023,Cammelli2025}. Pop III.1 stars are defined to be those that form from metal-free minihalos that are not impacted by external sources of feedback, while Pop III.2 stars form from metal-free minihalos that are affected by such feedback \citep{McKee2008}. Within the fiducial Pop III.1 SMBH seeding model, {\it all} Pop III.1 stars are assumed to undergo slow, monolithic collapse that adiabatically contracts a co-spatial dark matter cusp of the minihalo and thus are significantly impacted by WIMP self-annihilation heating to enable accretion of most of the baryonic mass of their halos \citep[][]{Nandal2025,Topalakis2025}. 
On the other hand, if a metal-free dark matter minihalo is impacted by a source of external EUV radiation, the abundance of free electrons in the gas increases, leading to enhanced formation of molecular hydrogen, higher cooling rates and a higher likelihood of fragmentation into multiple stars \citep{Greif2006,Johnson2006}, which would prevent adiabatic contraction of a dark matter cusp. While the validity of the assumptions of this Pop III.1 seeding model remain debated \citep[e.g.,][]{2025JCAP...11..009H}, it serves as the basis of a simple, physically-motivated scenario that can be implemented in cosmological simulations to yield testable predictions.

These above considerations explain why dark matter powered supermassive stars need to form from metal-free, pristine (i.e., negligible impact from external ionization) gas, i.e., Pop III.1 sources, and thus are predicted to be born isolated from each other. Our model uses this isolation criterion in order to seed Pop III.1 stars and subsequent SMBHs into a cosmological simulation of dark matter halos. The first minihalo to form in a certain isolated region becomes the host of a Pop III.1 star (see Fig.~\ref{fig:seeding_schem}). Its close neighbours (within the feedback distance of the Pop III.1 star) do not meet the criteria to form a supermassive star, but rather are assumed to form clusters of lower-mass Pop III.2 stars. Papers I \citep{Banik2019} and II \citep{Singh2023} developed this seeding model using a constant isolation distance, which was a free parameter, $d_{\rm iso}$, that was varied. In this work, we expand on the previous models by developing a self-consistent isolation criterion, which we reformulate as the requirement to be separated from previously formed sources by at least a feedback distance, which is based on the expected size of the HII regions surrounding Pop III.1 and Pop III.2 stars.

This paper is organized as follows. In \S\ref{sec:methods} we describe the methods, highlighting differences from the previous papers in this series. In \S\ref{sec:results} we show the results of a variety of models that demonstrate how each parameter influences the seeding outcome. In \S\ref{sec:discussion} we compare our fiducial model to observational data, and make first predictions for the expected dual AGN fraction as a function of redshift and observable SMBH merger rates. Finally, we present our conclusions in \S\ref{sec:conclusions}.

\section{Methods} \label{sec:methods}

\subsection{Simulation}
\label{sec:simulation}

As in previous papers in this series, we utilize \textsc{Pinocchio}, which is a Lagrangian Perturbation Theory \citep[LPT; e.g.,][]{Moutarde1991} code used for cosmological simulations \citep{Monaco2002,Munari2017}. The code follows the approximate evolution of density perturbations in a \lcdm{} universe, thus simulating the origin and evolution of dark matter halos. A \textsc{Pinocchio} run outputs catalogues of the halos at a set of pre-defined redshifts, containing their masses, positions and velocities. In addition to that, the complete merger history of each halo is stored with continuous time sampling.

The simulation used for this work is nearly identical to the one described in \citet{Singh2023}, with the only difference being the random seed used to generate the initial conditions. Here we give a brief summary of the setup, and we direct the reader to \S2.1 of \citet{Singh2023} for more details (see their high-resolution simulation). We have simulated a cosmological box with a side of 59.7~cMpc ($40~h^{-1}~\mathrm{cMpc}$, $h=0.67$) using standard Planck cosmology \citep{Planck2020}. The box was generated with V5 of \textsc{Pinocchio} 
on 800 MPI tasks (over 100 computing nodes with 256~GB of RAM per node), and it contains $4096^3$ particles of mass $1.23\times 10^5\:M_{\odot}$. As in Paper II, we have adopted a minimum mass of 10 particles for the halo identification,
giving us a minimum halo mass of $1.23\times 10^6\:M_{\odot}$. However, it is a known issue of \textsc{Pinocchio} that it underestimates the halo masses by a factor of $\sim 2$ at $z \sim 30$ (see \S2.1 of \citealt{Singh2023}). This does not affect the halo positions, but it means that our identified halos correspond to somewhat more massive sources than $10^6\:M_\odot$ and/or can be considered to be $10^6\:M_\odot$ halos after a modest amount of time evolution.

The SMBH seeding is performed on simulation data down to $z=10$, as the different seeding criteria explored here mainly affect the number density and distribution of the SMBHs at higher redshifts ($z \gtrsim 15$). Additionally, we follow halo mergers down to $z=0$ using a low-resolution version of the same simulation run, as described in \citet[][see their \S2.1]{Singh2023}. This allows us to take into account the decrease in SMBH comoving number density due to mergers. This switch to a low-resolution simulation for tracking the mergers was implemented in order to reduce computational costs.

\begin{figure}[ht!]
	\includegraphics[width=\linewidth]{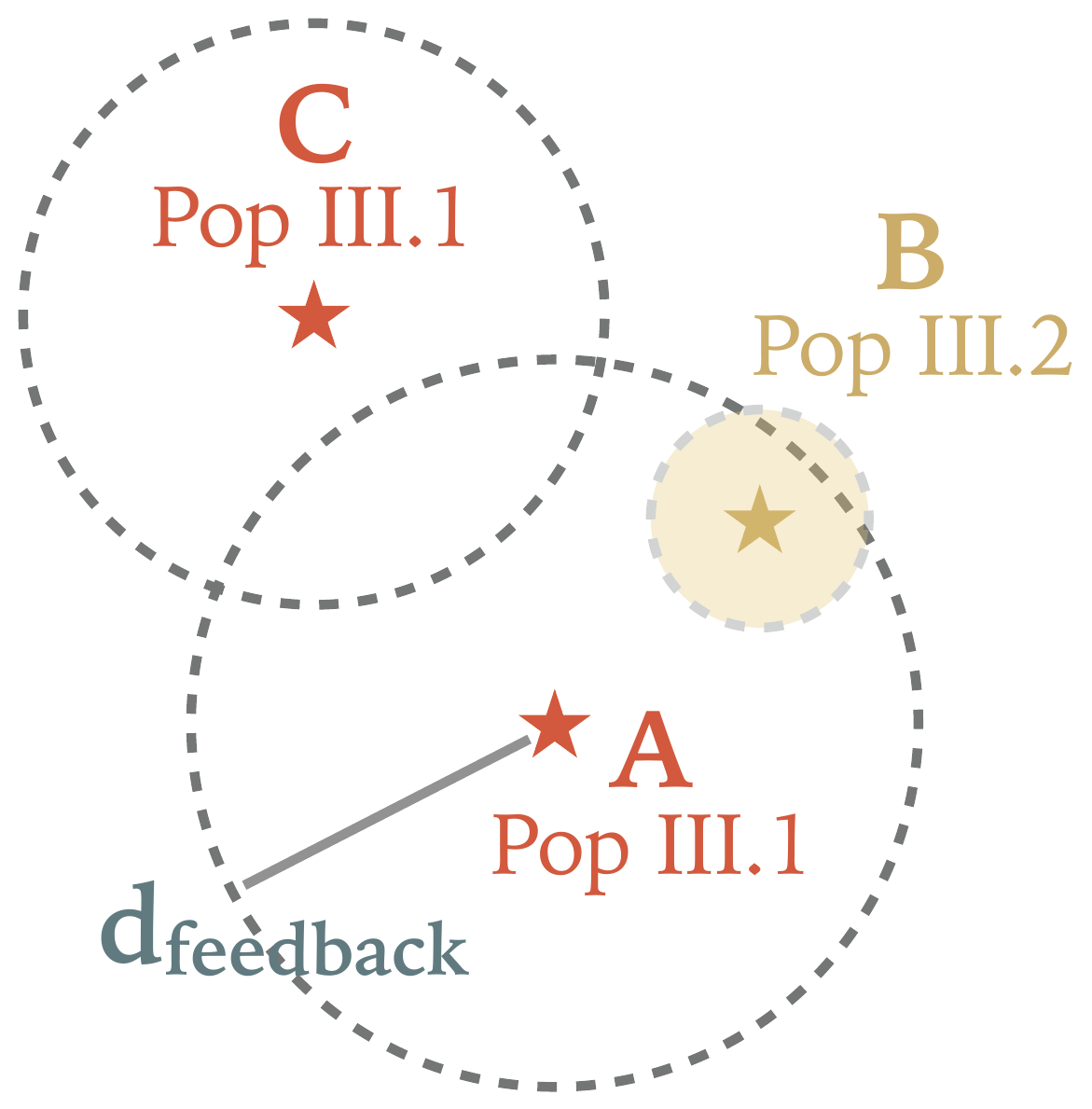}
    \caption{Schematic of Pop III.1 SMBH seeding model. 
    The first halo to appear in this region, A, is a Pop III.1 source and is seeded with a SMBH. Each subsequent halo is seeded as a Pop III.1 star / SMBH only if it falls outside the feedback range of a previously existing ionization source. Thus halo B remains unseeded, but halo C receives a seed. Unlike in previous work \citep[][]{Banik2019,Singh2023}, here each Pop III.1 star has its individual feedback distance, and in some of the models the unseeded (Pop III.2) halos are also included as more modest ionization sources.}
    \label{fig:seeding_schem}
\end{figure}

\subsection{SMBH seeding algorithm} \label{sec:seeding}

The seeding of Pop III.1 stars and subsequent SMBHs is performed as a post-processing step to the output of the \textsc{Pinocchio} simulation. Here we use a method very similar to the one described in \S2.3 of \citet{Singh2023} with a few key modifications. In this section, we provide a summary of the seeding scheme of \citet{Singh2023}, highlighting the differences with this work. 

The SMBH seeding algorithm visits all halos in the order of their appearance and determines whether or not they should be populated with Pop III.1 / SMBH sources. 
The first identified halo within the box is automatically seeded, and the algorithm ensures that any subsequently seeded halos are outside the feedback range of existing ionizing sources (see Figure~\ref{fig:seeding_schem}). We assume that all seeds are (for some amount of time) ionizing sources, but a key difference with the previous work is that here each seed has its own individual feedback distance which evolves with time (discussed in more detail in \S\ref{sec:feedback}), while previously the feedback distance was a free parameter that was kept constant in proper distance units for all halos at all times. In addition to the seeded sources, in some of our models we also consider the ionizing feedback from non-seeded halos, i.e., Pop III.2 sources, which can have their own feedback distances (see \S\ref{sec:feedback}). In contrast, \citet{Singh2023} did not distinguish between seeded and non-seeded halos when they applied their separation criteria.

The execution of this algorithm is facilitated by the following technical steps to achieve computational efficiency. The algorithm first splits the time domain into redshift bins and visits these in order of time evolution. In our analysis we used $z=\{37, 32, 30, 29, 26, 25, 23, 20, 18, 16, 15, 14, 12, 10\}$ as bin boundaries. The upper limit bin boundary of $z=37$ was chosen as the first minihalo in this simulation appears around $z=40$. The halos that exist within a given redshift are selected and their positions at the higher redshift boundary of the bin are used for building a kd-tree. Note that the tree typically includes a significant number of halos that are only present for a part of the redshift bin, i.e., they either appear during the timespan of the bin and/or merge with larger halos and are therefore removed from the list of halos. 

The halos that appear in the current bin are then visited in their order of appearance. As a newly appeared halo is considered, we find its neighbours using the kd-tree\footnote{For some of the models presented here we have used an alternative version of the algorithm where the kd-tree is built only from halos that are already seeded by the starting redshift of the bin. This ``seed'' tree is typically significantly smaller and faster to search, however it needs to be supplemented by additional (brute-force) checks of the halos that are seeded \textit{during} the current redshift bin since they would be missing from the tree. We have confirmed that the two implementations of the kd-tree produce identical results.}. The search distance that is chosen is the maximum of the existing feedback distances of already seeded halos plus a small tolerance that accounts for the small peculiar motions of the halos during the redshift bin. This was chosen due to the fact that it is too computationally expensive to reconstruct the kd-tree each time that a new halo appears. The same step was performed in \citet{Singh2023}, but with a constant search radius.

Once the neighbours of the new halo are found, the algorithm checks if any of them are currently present in the simulation (discarding ones that appear later in the bin and ones that have merged) and have already been seeded. In the runs where only feedback contributions from Pop III.1 stars are considered, we discard the non-seeded halos. We then compute the distances from the identified seeded halos to the new halo, and compare them to the corresponding individual feedback distances. The new halo is seeded if it lies outside the feedback distance of all other seeds. In the runs that also include Pop III.2 sources, we similarly apply an additional distance criterion. Upon seeding, the new halo is assigned its own individual feedback distance (see the rest of this section), which is propagated together with the seed itself to a new halo when a merger occurs. 
 
\subsection{Feedback distance}
\label{sec:feedback}

An important aspect of our model is that Pop. III.1 stars may only form from pristine, non-irradiated gas in order to avoid fragmentation, enabling significant adiabatic contraction of the dark matter density during star formation, and thus reach sufficiently high stellar masses. The implication of this condition is that Pop. III.1 stars need to form in isolation from each other's ionization feedback. Therefore within this work we define a feedback distance that we use as a separation criterion in the seeding algorithm (see \S\ref{sec:seeding}).

The previous papers of this series treated the feedback distance (\diso{}) as a free parameter, which was kept as a constant proper distance within the simulation. In order to place the current work in the context of previous results, we first expand this parameterization of the feedback distance by considering a set of models where \diso{} is treated as constant in \textit{comoving} units. As demonstrated later, the constant comoving \diso{} models are a good representation of our preferred physical models. In addition, we also develop models which compute the size of an H~\textsc{II} region as a function of redshift and assumed Pop III.1 stellar properties (see Table~\ref{tab:models} for a complete list of models).

The characteristic size of the HII region around a massive star is typically given in terms of its Str\"omgren radius \citep{Stromgren1939}. This is an equilibrium solution where the rate of emission of ionising photons from the star balances out the recombination rate of the ionised gas to a non-ground energy level\footnote{If an electron is recaptured directly to its ground state, the photon that is emitted in the recombination event will be able to ionise another hydrogen atom, and hence does not lead to a net ``loss'' of ionising flux.}. For a star with hydrogen ionizing photon rate of $S$ in an ambient medium of average density $n_{\rm H}$, the Str\"omgren radius is given as:
\begin{equation}
    R_{\rm S} \approx 59.6\:S_{53}^{1/3} \left(\frac{T}{ 3 \times 10^4~\rm K}\right)^{0.27} \left(\frac{n_{\rm H}}{n_{\rm H,z=30}} \right)^{-2/3}{\rm kpc}.
    \label{eq:stromgren}
\end{equation}
In the equation above, $T$ is the temperature of the ionised gas (here assumed to be pure hydrogen), $n_{\rm H,z=30}$ is the mean baryonic density at $z=30$, and $S_{53}\equiv S/10^{53}~\rm s^{-1}$.

The timescale on which the size of the H II region reaches the Str\"omgren radius is given by the recombination time:
\begin{equation}
    t_{\rm rec} = \frac{1}{\alpha_{\rm B}n_{\rm H}} \approx 20.6~{\rm Myr} \left(\frac{z+1}{31} \right)^{-3},
    \label{eq:rec-time}
\end{equation}
where $\alpha_{\rm B}$ is the recombination coefficient to non-ground levels. 

Equation~\ref{eq:rec-time} shows that for a typical formation redshift of 30, the recombination time is likely to exceed the lifetime of a high-mass star, and therefore the Str\"omgren radius might not be reachable. Since the (high ionization rate portion of the) lifetime of Pop III.1 stars is not well constrained, e.g., it may be prolonged compared to the standard main sequence lifetime of $\sim 3\:$Myr due to residual WIMP annihilation heating, we consider it to be a free parameter in our model. For a source emitting for a time $t<t_{\rm rec}$, the size of the H II region is approximately given by the R-type expansion equation \citep{Kahn1954}:
\begin{equation}
    R_{\rm R} = R_{\rm S} \left(1 - \exp\left(-t/t_{\rm rec}\right)\right)^{1/3}
    \label{eq:r-type}.
\end{equation}

\begin{figure}
	\includegraphics[width=\linewidth]{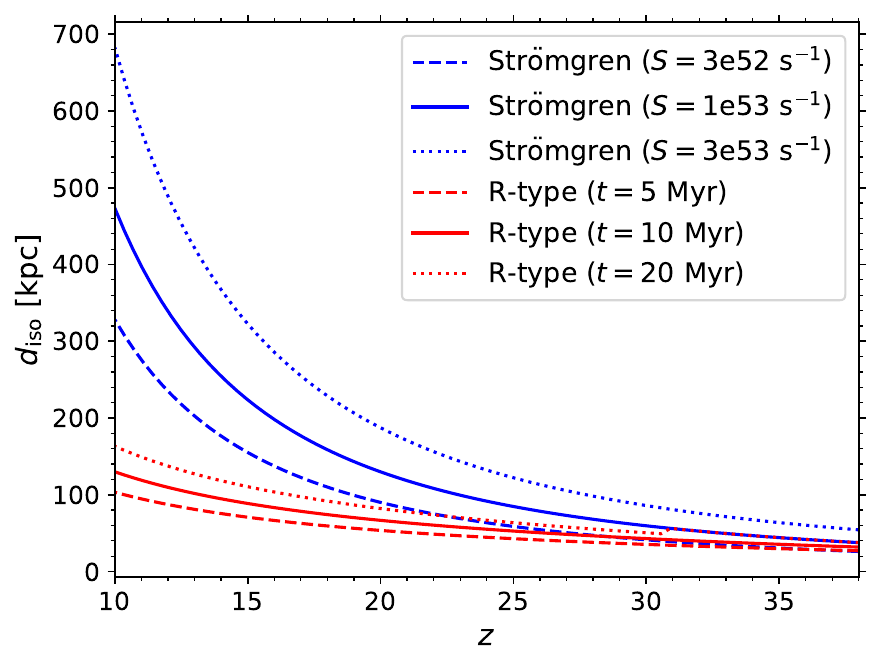}
    \includegraphics[width=\linewidth]{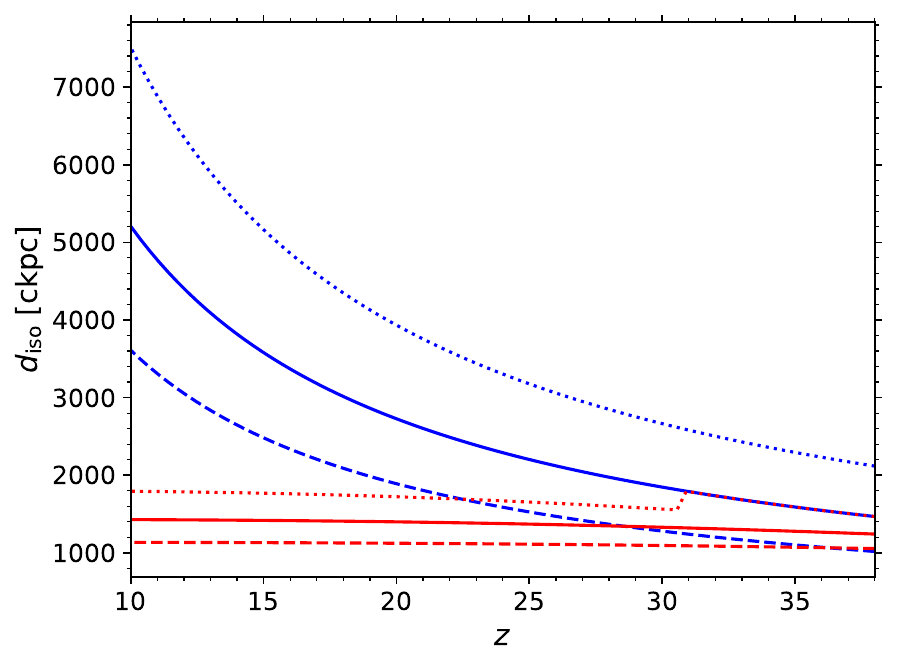}
    \caption{Isolation distance as a function of seeding redshift in proper (\textit{top}) and co-moving coordinates (\textit{bottom}). The blue lines show the Str\"omgren radius and follow eq.~\ref{eq:stromgren}, assuming a range of values for $S$ and $T=3\times 10^4~\rm K$. The red lines follow eq.~\ref{eq:r-type}, assuming a range of stellar lifetimes of $t=10~\rm Myr$, and the value for $R_{\rm S}$ corresponding to $S=10^{51}~\rm s^{-1}$. Note that the jump in the red dotted line comes from the fact that when the lifetime of the star exceeds the recombination time (given by eq.~\ref{eq:rec-time}), eq.~\ref{eq:r-type} is replaced by eq.~\ref{eq:stromgren}.}
    \label{fig:diso-vz-z}
\end{figure}

In order to explore a broad range of possible feedback distances, we run models where the Pop III.1 stars are assumed to have long lifetimes (using the Str\"omgren radius given in eq.~\ref{eq:stromgren}), and also ones with shorter-lived Pop III.1 stars (governed by the R-type expansion in eq.~\ref{eq:r-type}). The two parameters that we vary are $S$ ($10^{52.5},~10^{53},~10^{53.5}~\rm s^{-1}$) and $t$ ($5,~10,~20~\rm Myr$), with our fiducial model being $S=10^{53}~\rm s^{-1}$ and $t=10~\rm Myr$ 
(see Table~\ref{tab:models}). Note that within the shorter-lived models whenever $t>t_{\rm rec}$ is encountered, we use the Str\"omgren radius instead of the R-type radius. The feedback distances resulting from these parameter choices are shown in Fig.~\ref{fig:diso-vz-z}. The bottom panel of the figure shows that while the Str\"omgren radius varies in both proper and co-moving units as a function of redshift, the R-type radius remains approximately constant in co-moving units. The constant co-moving size is not an artifact of the specific parameter choices, and can also be shown analytically by Taylor expanding the exponential term of eq.~\ref{eq:r-type}, and combining it with eq.~\ref{eq:stromgren} and \ref{eq:rec-time} \citep{Tan2024,Sanati2025}:
\begin{eqnarray}
    R_{\rm R} \approx 1.45\:S_{53}^{1/3} \left(\frac{T}{ 3 \times 10^4\rm K}\right)^{0.27} \left(\frac{t}{10\rm Myr}\right)^{1/3}{\rm cMpc}
    \label{eq:r-type-approx}.
\end{eqnarray}

\begin{table*}
	\centering
    \caption{Summary of model parameters. Where ranges of values are given, they capture the range for $z=10-40$.} 
	\label{tab:models}
	\begin{tabular}{lccccccr} 
		\hline
         &  & Pop. III.1 & & & & Pop. III.2 &\\
        Model name & $\log S~[\rm s^{-1}]$ & $t$ [Myr]& $t_{\rm del}$ [Myr] & \diso{} [cMpc] &    & $\log S_2~[\rm s^{-1}]$ & $d_{\rm iso,2}$ [cMpc]\\
		\hline
        CoMov-all-1 & -& -& 0  & $1.0$ & & - & $1.0$\\
        CoMov-all-2 & -& -& 0  & $2.0$ & & - & $2.0$\\
        CoMov-all-3 & -& -& 0  & $3.0$ & & - & $3.0$\\
        CoMov-1 & -& -& 0  & $1.0$ & & - & -\\
        CoMov-2 & -& -& 0  & $2.0$ & & - & -\\
        CoMov-3 & -& -& 0  & $3.0$ & & - & -\\
        Str-52.5 & $52.5$& -& 0  & $1.0-3.5$ & & - & -\\
        Str-53 & $53$& -& 0 & $1.5-5.2$ & & - & -\\
        Str-53.5 & $53.5$& -& 0 & $2.1-7.5$ & & - & -\\
        R-5 & $53$& 5 & 0 & $\sim 1.1$ & & - & -\\
        R-10 & $53$& 10 & 0 & $\sim 1.3$ & & - & -\\
        R-20 & $53$& 20 & 0 & $\sim 1.6$ & & - & -\\
        RDel-50& $53$& 10 & 50 & $\sim 1.3$ & & - & -\\
        RDel-100& $53$& 10 & 100& $\sim 1.3$ & & - & -\\
        Fiducial & $53$& 10 & 50 & $\sim 1.3$ & & 51 & $0.3-1.1$ \\
		\hline
	\end{tabular}
\end{table*} 

Unlike the models presented in \citet{Banik2019} and \citet{Singh2023}, we consider the fact that the relic H II regions will grow in proper size even after the star is gone due to the expansion of the universe. For that reason, the feedback distance that each seed is assigned is assumed to remain constant in co-moving units. 

In addition to the ionizing luminosity from the Pop III.1 sources, we also consider in some of our models the ionization caused by high-mass stars contained within the non-seeded mini-halos. Since we do not model star formation explicitly, we assume that a number of such high-mass stars would be present at all times after the formation of the halo, and therefore we use the associated Str\"omgren radius as the feedback distance. We assign the non-seeded sources $S_2=10^{51}~\rm s^{-1}$. For reference, that would be a luminosity typical for a stellar cluster of mass $2\times 10^4~\rm M_{\odot}$ if we were to assume a well sampled IMF at $z=0$ \citep{Leitherer2014}. While the exact value of the luminosity is not known for these high redshift objects, the fact that the universe is not yet fully reionized before $z\sim 20$ suggests that their ionising flux contribution would be modest. In addition, since we assume a steady self-replenishing population of high-mass stars in these sources, we do not consider a relic H II region that is expanding with $(z+1)^{-1}$, as we have with the Pop. III.1 sources. Instead, we compute their instantaneous Str\"omgren radius at each redshift, which grows faster ($\propto (z+1)^{-2}$; see eq.~\ref{eq:stromgren}).

\subsection{Formation time delay}
\label{sec:delay}

The previous papers in this series assumed that the Pop III.1 stars would appear at the same time as the mini-halo forms within the simulation. However, more realistically we expect a time delay between these two events, as the formation of the star would require the baryonic matter to condense to the central part of the mini-halo. 
A way to estimate this delay is by considering the free-fall time. At a typical formation redshift of 30, the average IGM number density of Hydrogen nuclei is $\sim 5.72\times 10^3~\rm cm^{-3}$. If the mini-halo contains an overdensity of baryonic matter by a factor of $10-100$ compared to the average, the free-fall time ($t_{\rm ff}=\sqrt{3\pi/32G\rho}$) is approximately 50-160~Myr. Such delays between the halo formation/identification time and the onset of star formation are also seen in hydrodynamical cosmological simulations, and are also typically of the order of 50-100~Myr \citep[e.g.,][]{Sanati2025}, however the exact time is sensitive to the resolution used in the simulation. Given these considerations, we also consider the possibility of a time delay from formation of a Pop III.1 minihalo to its star.

The delay is taken into account within the seeding algorithm as follows. We apply a constant time delay for all Pop. III.1 stars, given by the free parameter $t_{\rm del}$. Since the formation of all stars is delayed by the same amount, the order of seeding remains unchanged. This allows us to keep visiting the mini-halos in their order of appearance. The main differences come in the feedback distance, which is calculated using a time-delayed redshift, and in the feedback distance criterion being applied at a delayed redshift. To sample a wide range of plausible values for $t_{\rm del}$, we have considered models with a delay of 0, 50 and 100~Myr (see Table~\ref{tab:models}), while choosing $t_{\rm del} = 50$~Myr as our fiducial model.

\subsection{Mergers}
\label{sec:mergers}

A direct implication of our model is that the SMBH seeds are initially distant from each other and unclustered. However, at lower redshift some halos might contain multiple seeds due to halo mergers. Since we track the propagation of each seed from one halo to another via mergers, we also record the redshift at which an already seeded halo acquires a new seed. We refer to these events as `seed mergers' for simplicity, even though in reality there should be a time delay between the merging of dark matter halos and the SMBHs that they contain.

The seeding of the \textsc{Pinocchio} simulation is performed down to $z=10$ (see Sec.~\ref{sec:seeding}), and we can track the seed mergers until $z=8$. For all of our models, only very few seeds merge so early ($\lesssim 0.05\%$ of all seeds), which has a negligible effect on our results. For our fiducial model, we also follow the seed mergers to $z=0$. Since extending our high resolution run for that long has a prohibitively high computational cost, we perform a low resolution version of the simulation (with $1024^3$ particles) for the purposes of tracking the seeds. This is analogous to the procedure described in Sec.~2.1 of \citet{Singh2023}. Instead of tracking the low resolution mergers only from $z=8$ onwards, we go through the full merger trees of the low resolution simulation (since $z\sim30$) and identify seed mergers independently.

Note that \textsc{Pinocchio} considers two halos to merge when their separation becomes smaller than $D=0.35 R_{\rm L,1}$, where $R_{\rm L,1}$ is the Lagrangian radius of the more massive halo, and it depends on the mass of the halo, $M_1$. \citet{Monaco2002} express this merger distance as 
\begin{equation}
    D=0.35 R_{\rm L,1} = 0.35 \left (\frac{3M_1}{4\pi\Omega_M\rho_c} \right)^\frac{1}{3}.
    \label{eq:merg-dist}
\end{equation}
Since a lower resolution simulation can only resolve the higher mass halos, the minimum separation of a merger event is also larger. Therefore, the definition of a merger event is not consistent between the high resolution and the low resolution runs, and we will only consider the low resolution mergers for our full merger analysis.



\section{Results}
\label{sec:results}

\subsection{Constant co-moving \diso{}}

\begin{figure}
    \includegraphics[width=\linewidth]{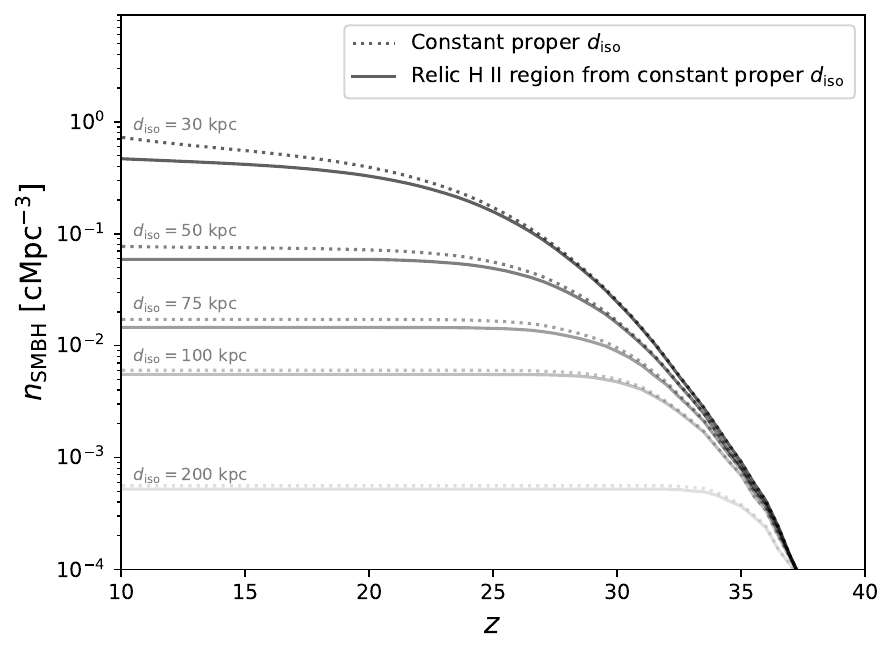}
    \caption{Co-moving number density of SMBH seeds as a function of redshift. The grey dotted lines show seeding models with constant \diso{} \citep[see][]{Banik2019,Singh2023}, while the solid lines show models with \diso{} that is constant upon seeding, but is subsequently scaled by a factor $(1+z)^{-1}$.}
    \label{fig:num_dens_const}
\end{figure}

\begin{figure}
    \includegraphics[width=\linewidth]{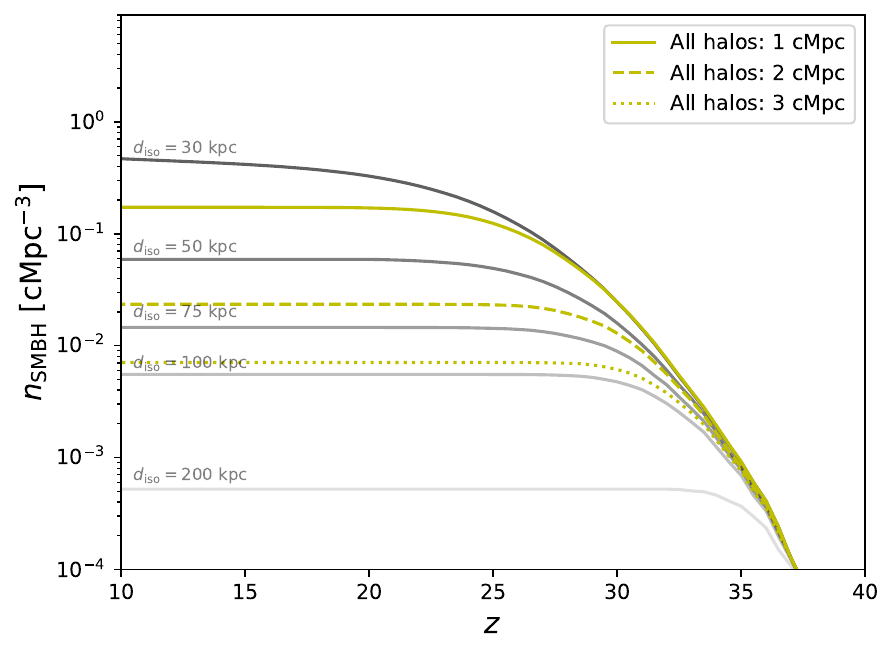}
    \includegraphics[width=\linewidth]{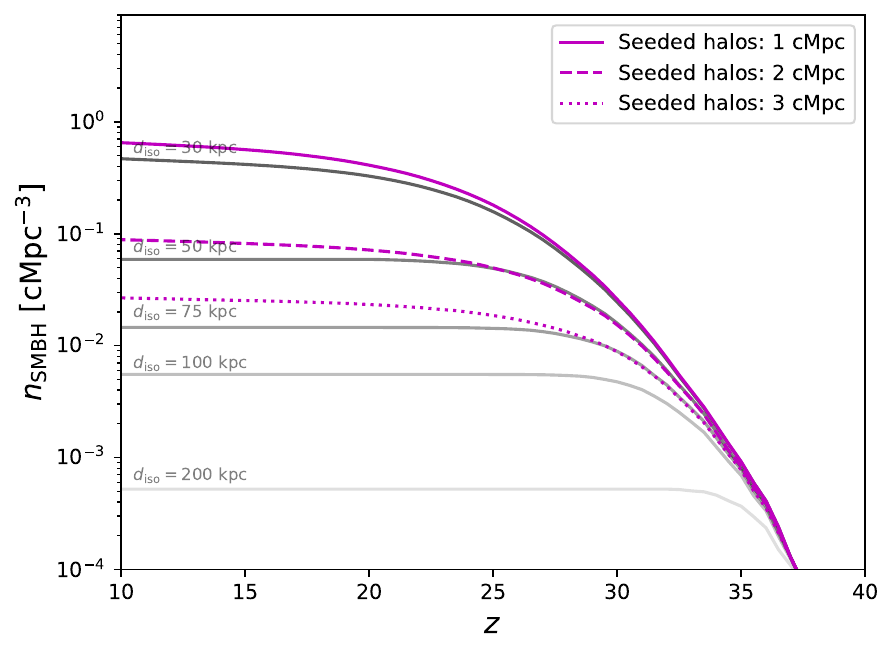}
    \includegraphics[width=\linewidth]{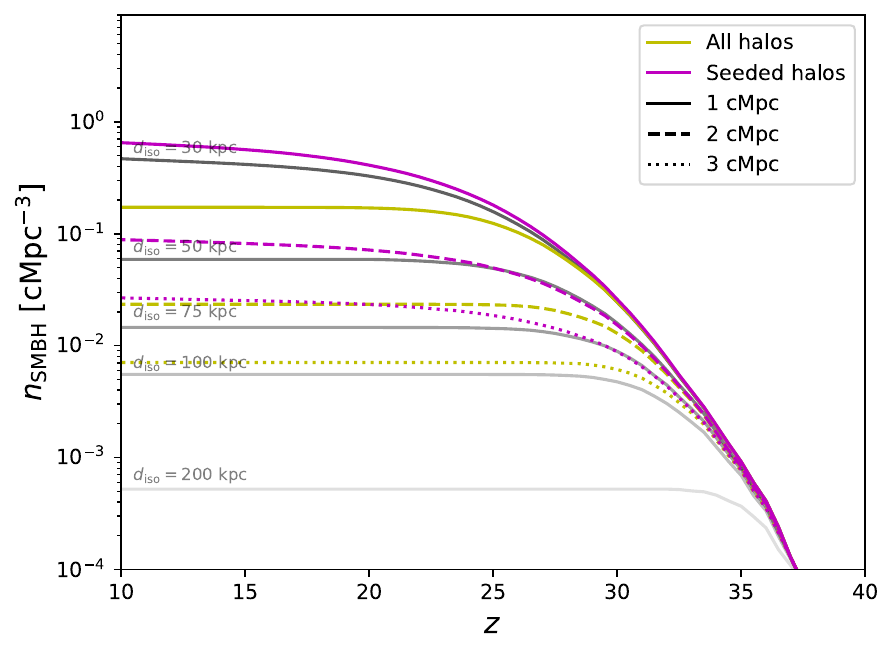}
    \caption{Co-moving number density of SMBH seeds as a function of redshift. The grey solid lines show models with \diso{} that is constant upon seeding, but is subsequently scaled by $(z+1)^{-1}$ (as in Figure~\ref{fig:num_dens_const}). The coloured lines show models where \diso{} is constant in co-moving units upon seeding and is subsequently kept constant in co-moving units (this is equivalent to the $(z+1)^{-1}$ scaling of the grey lines). The distance criterion in the models was applied around all halos (yellow) or only around the seeded halos (magenta).}
    \label{fig:num_dens_const_comov}
\end{figure}

Since the goal of this work is to construct more physically-motivated, feedback-regulated models of the isolation distance, the first change that we implement from the previously published models of this paper series is to consider that a relic HII region in the early Universe would continue to expand with the Hubble flow. We consider this to be a reasonable approximation, since the typical H II region size is much larger than the radius of the Pop III.1 halo \citep[see also][]{Sanati2025}.

Figure~\ref{fig:num_dens_const} shows a comparison between the previous models presented in \citet{Singh2023}, and the ones where \diso{} is constant in \textit{proper} units upon seeding, but is subsequently scaled by $(z+1)^{-1}$ when the distance checks for future seeding are performed. Note that in both cases the distance criterion is applied from all existing halos, regardless if they are seeded or not. The scaling results in an increase in the isolation distance as time progresses, and hence more of the simulation volume is filled with regions where subsequent seeding would be prohibited (compared to the approach in \citealt{Singh2023}). Therefore, the scaled \diso{} models produce modestly lower comoving SMBH number densities (by $\sim 5-25\%$). Note that all models start off with a period of rapid seeding and hence an increase of the co-moving number density of SMBHs, and then reach a plateau when no further seeding is happening.

\begin{figure*}
    \includegraphics[width=\linewidth]{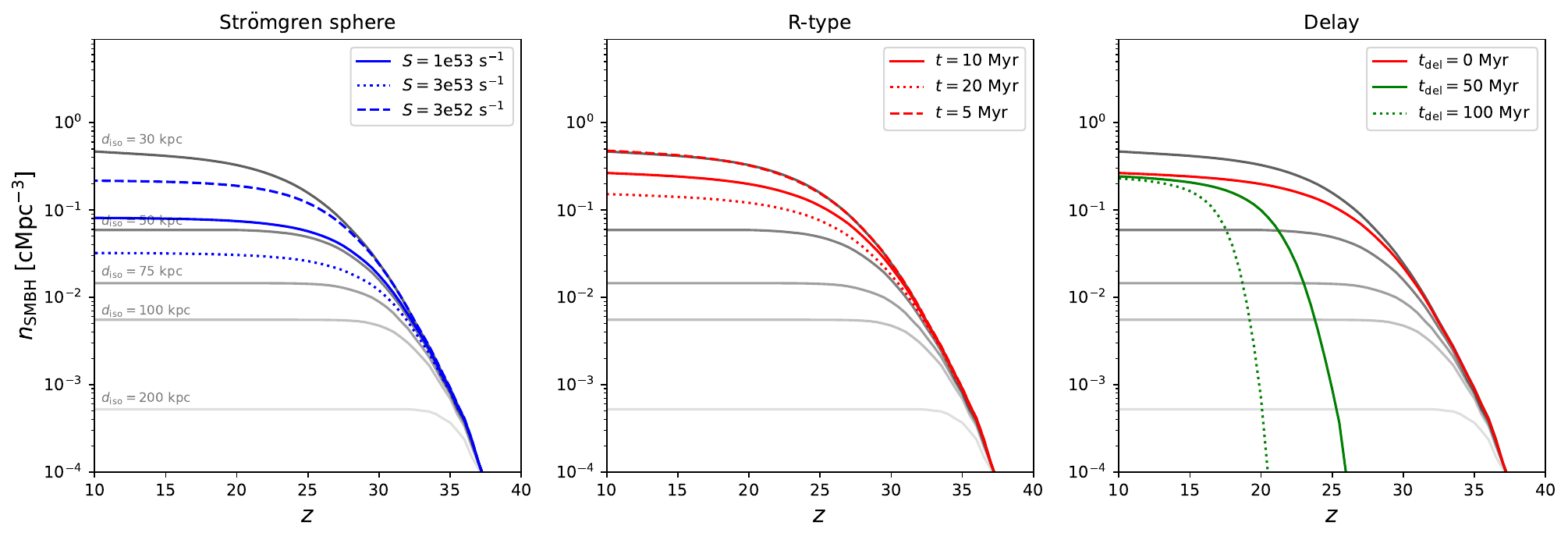}
    \caption{Co-moving number density of SMBH seeds as a function of redshift. The grey lines show seeding models with constant seeding isolation distance that is scaled by $(z+1)^{-1}$ (see Figure~\ref{fig:num_dens_const}), while the coloured lines show models with feedback-regulated isolation distances.}
    \label{fig:num_dens_all_models}
\end{figure*}

\begin{figure*}
    \includegraphics[width=\linewidth]{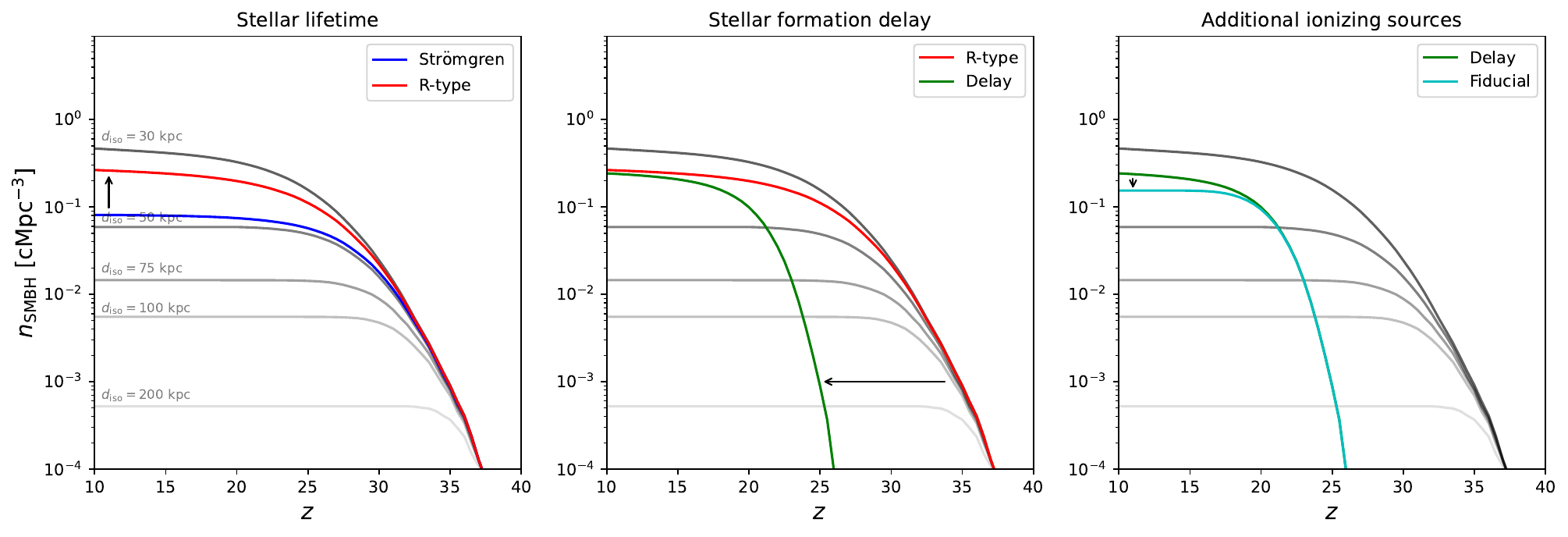}
    \caption{Same as Figure~\ref{fig:num_dens_all_models}, but showing select models that demonstrate the build up to the fiducial one. We begin with a Str\"omgren radius model (Str-53) and include progressively the effect of finite stellar lifetime (R-10) and the formation delay of the Pop.~III.1 sources (RDel-50), as well as the impact of Pop.~III.2 sources (Fiducial).}
    \label{fig:num_dens_path_to_fid}
\end{figure*}

We then consider models with \diso{} that is constant in co-moving units upon seeding and subsequently when the distance criterion is applied. This is similar to the models scaled by $(z+1)^{-1}$, but the upon-seeding value of \diso{} is now different in proper units as a function of redshift. The models with constant co-moving \diso{} are shown in Figure~\ref{fig:num_dens_const_comov}. The top panel of the figure shows models where the distance criterion is applied around all halos (named  CoMov-all-1, CoMov-all-2 and CoMov-all-3 in Table~\ref{tab:models}), and we see that the overall shape of the co-moving number density curves is approximately the same as in the models in Figure~\ref{fig:num_dens_const}, even though there is no direct correspondence between models. However, the shape of the curves changes when the distance criterion is applied only around the seeded halos (middle panel of Figure~\ref{fig:num_dens_const_comov}; models CoMov-1, CoMov-2 and CoMov-3 in Table~\ref{tab:models}). In these models the co-moving number density of seeds does not reach a plateau, but continues to increase even down to $z=10$. This can be explained by the fact that the distance criterion can be thought of as filling the simulation volume with spheres, and a new seed can be added only if its halo appears in the gap left between existing spheres. Therefore, the final stages of seeding (close to the upper limit imposed by geometric considerations) can only happen when a large enough number of new halos have appeared to place the final seeds in small existing gaps, which occurs at low redshifts. The close packing limit for models CoMov-1, CoMov-2 and CoMov-3 gives maximum co-moving number densities of 1.4, 0.18 and 0.052~cMpc$^{-3}$ respectively, which are approached but not reached by any of the models. In addition, the models with distance criterion only around seeds produce more seeded halos than their counterparts with distance criterion around all halos, as expected.


\subsection{Str\"omgren radius}
The left panel of Figure~\ref{fig:num_dens_all_models} shows the models that use a Str\"omgren radius (named Str-52.5, Str-53 and Str-53.5 in Table~\ref{tab:models}). The corresponding \diso{} values for these models were previously shown in Figure~\ref{fig:diso-vz-z}. The left panel of Figure~\ref{fig:num_dens_all_models} demonstrates that by varying the stellar ionizing luminosity by 0.5 dex from our fiducial value of $S=10^{53}~\rm s^{-1}$, we obtain a wide range of final co-moving SMBH densities ($\sim 0.03-0.2~\rm Mpc^{-3}$). These values fall within the $30-75$~kpc range for constant \diso{}. Note that the curves rise steeply until $z\sim 25-30$, and subsequently almost flatten by $z=10$, which indicates that most SMBH seeds are already in place early on in the simulation.

\subsection{Stellar lifetimes}

The middle panel of Figure~\ref{fig:num_dens_all_models} shows the models that use \diso{} corresponding to Pop III.1 stars with various lifetimes (named R-5, R-10 and R-20 in Table~\ref{tab:models}). All of these models assume our fiducial value for the stellar luminosity of $S=10^{53}~\rm s^{-1}$, and the only parameter that is varied is the stellar lifetime (see Figure~\ref{fig:diso-vz-z} for the corresponding \diso{} values). By choosing stellar lifetimes that are shorter than the timescale on which the Str\"omgren radius of the H II region is established (see eq.~\ref{eq:rec-time}), results in smaller values for \diso{} and an increase of the final SMBH co-moving number density by a factor of a few ($\sim 0.2-0.5~\rm Mpc^{-3}$) compared to the corresponding Str\"omgren case ($\sim 0.08~\rm Mpc^{-3}$). These models fall within the $30-50$~kpc range for constant \diso{}. They also keep increasing in terms of co-moving number density for longer than the Str\"omgren case, and do not fully manage to reach their plateau by $z=10$. 

\subsection{Formation delay}

The right panel of Figure~\ref{fig:num_dens_all_models} shows the models that include a time delay between the halo formation and Pop III.1 formation (i.e., the moment of seeding). For these models we assume stellar luminosity of $S=10^{53}~\rm s^{-1}$, and a stellar lifetime of $t=10~\rm Myr$, but we vary the delay time (models RDel-50 and RDel-100 in Table~\ref{tab:models}). For comparison we also show the corresponding model with no delay (same as the solid red line in the middle panel of Figure~\ref{fig:num_dens_all_models}). The models presented in this panel have different starting seeding redshifts, however, they all have most of their seeding completed by $z=10$. Furthermore, they reach approximately the same final co-moving SMBH number density ($\sim 0.3~\rm Mpc^{-3}$). This is due to the fact that the models with short stellar lifetimes (capturing the R-type expansion of the H II regions) reach approximately constant H II region sizes in co-moving units (see the bottom panel of Figure~\ref{fig:diso-vz-z}), and hence the seeding redshift does not affect the size of \diso{} in comparison to the simulation box. 

\subsection{Pop III.2 ionizing sources}

\begin{figure}
    \includegraphics[width=\linewidth]{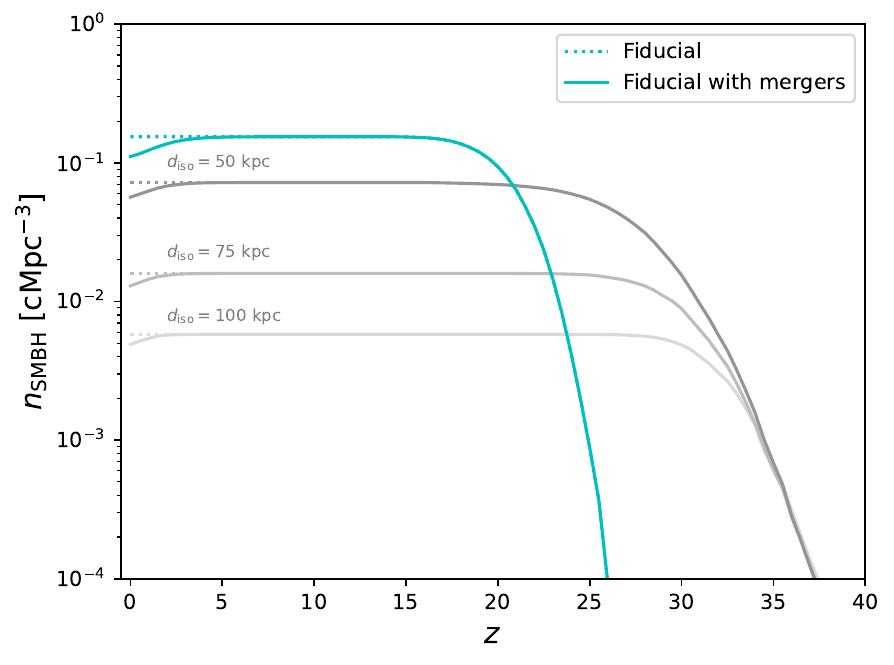}
    \caption{Co-moving number density of SMBH seeds as a function of redshift calculated to $z=0$. The grey lines show seeding models from \citet{Singh2023}, while the cyan line shows the fiducial model.}
    \label{fig:num_dens_fiducial}
\end{figure}

Finally, Figures~\ref{fig:num_dens_path_to_fid} and~\ref{fig:num_dens_fiducial} shows our fiducial model, which assumes a Pop III.1 luminosity of $S=10^{53}~\rm s^{-1}$, a stellar lifetime of $t=10~\rm Myr$, and a seeding delay of $t_{\rm del}=50~\rm Myr$ (see Table~\ref{tab:models}). In addition to the Pop III.1 sources, this model also includes distancing criteria from the non-seeded sources. These additional ionization sources slightly decrease the final co-moving number density of SMBHs ($0.16~\rm Mpc^{-3}$) compared to the corresponding model with only Pop III.1 sources (see right panel of Figure~\ref{fig:num_dens_path_to_fid}). This number density falls within the $30-50$~kpc range for constant \diso{}. Within the fiducial model, all seeded mini-halos are already in place by $z \sim 15-20$.

Figure~\ref{fig:num_dens_fiducial} extrapolates the co-moving number density prediction of the fiducial model to $z=0$ (see the cyan dotted line), and we also include a version of the curve in which it is reduced by the ongoing merger events that occur at lower redshifts (solid cyan line). This assumes that the number of SMBHs is reduced by one at the moment when two seeded halos merge. The mergers are tracked using the low resolution version of the simulation, as described in Section~\ref{sec:mergers}. For comparison, Figure~\ref{fig:num_dens_fiducial} includes the constant proper \diso{} models of \citet{Singh2023}, where effects of mergers are also included. We see that our fiducial model follows a very similar decrease in co-moving number density at $z\lesssim5$ to the constant proper \diso{} models.


\section{Discussion}
\label{sec:discussion}

\begin{figure*}
    \includegraphics[width=\linewidth]{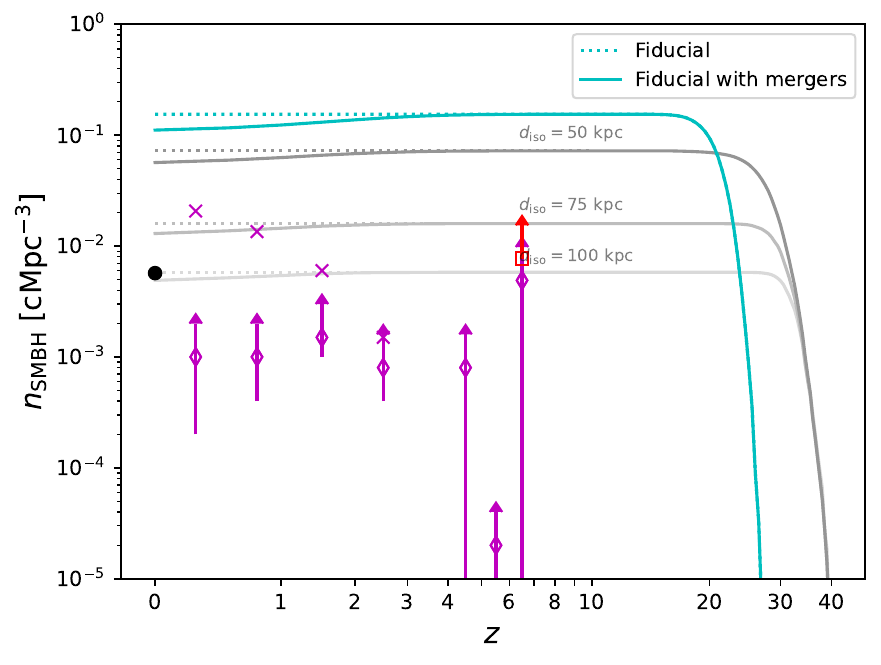}
    \caption{Co-moving number density of SMBH seeds as a function of redshift calculated to $z=0$. The grey lines show seeding models from \citet{Singh2023}, while the cyan line shows the fiducial model. The symbols show number density estimates from various observational studies. The black circle is derived by integrating the black hole mass function of \citet{Shankar2020} down to $M_{\rm SMBH} = 10^5\: M_{\odot}$ (assuming a constant value in the range $10^5 - 10^6\:M_{\odot}$). The magenta crosses and diamonds show the variability-corrected and the luminosity-corrected estimates respectively from \citet{Cammelli2025b}, and the red square is an estimate by \citet{Hayes2024}.}
    \label{fig:num_dens_fiducial_data}
\end{figure*}

\subsection{Analytic estimate of SMBH abundance}

Since many of the models have \diso{} that is approximately constant in co-moving units, which is mainly set by the near constant R-type HII region radius from the Pop III.1 supermassive stars, we can compute the close packing limit ($n_{\rm SMBH} = \sqrt{2}/R_{R}^3$) which gives co-moving number densities of 1.1, 0.64 and 0.35~cMpc$^{-3}$ for R-5, R-10 (i.e., with $R_R=1.3\:{\rm cMpc}$) and R-20, respectively. The close packing limit indicates the maximum obtainable number density with an optimal arrangement of the seeds, and in practice it is not reachable by halos emerging via a more realistic distribution reflecting cosmic large scale structure. Indeed, we see that the close-packing limit values are a factor of $2-3$ higher than the obtained $z=10$ number densities.
We then describe the number densities reached by our models by $z=10$ via:
\begin{equation}
    n_{\rm SMBH}\simeq \frac{3\phi_V}{4\pi R_{\rm R}^3},
\end{equation}
where $\phi_V\approx2.4-3.0$ is a dimensionless factor that is calibrated from our numerical results, with $\phi_V\approx2.4$ for R-10 which most closely resembles the fiducial model.

\subsection{Comparison to observations and other models}

Our models predict that all SMBH seeds are in place by $z\gtrsim15$, and we obtain a broad range of co-moving SMBH number densities ($\sim 0.03-0.7~\rm cMpc^{-3}$), depending on the choice of parameter values. The high seeding redshifts easily accommodate for the recent observations of AGN at $z\gtrsim7$, which is a strong point in favour of this theoretical model. In addition to the existence of individual high redshift AGN, we compare our models to observational number density estimates (see Figure~\ref{fig:num_dens_fiducial_data}). The $z=0$ data point is derived by integrating the SMBH mass function of \citet{Shankar2020} down to $M_{\rm BH} = 10^5\:M_{\odot}$ (assuming a constant value in the range $10^5 - 10^6\:M_{\odot}$). The higher redshift data are deduced from a search for AGN via variability in the Hubble Ultra Deep Field (HUDF) \citep{Hayes2024,Cammelli2025b}. The number densities that we obtain with our models are around 1-2 orders of magnitude higher then the observed constraints, with our fiducial model being a factor of $\sim 30$ higher than the $z=0$ data point (see Figure~\ref{fig:num_dens_fiducial_data}). 
Therefore, a prediction of our seeding scheme is that there are many SMBHs that are present in the Universe, but that are missed observationally (due to their inactivity or due to sensitivity limits). 

The feedback-regulated \diso{} models presented here fall within the range of $30-75~\rm kpc$ of the constant \diso{} models used in \citet{Banik2019,Singh2023} and \citet{Cammelli2025}, with our fiducial model being between $30-50~\rm kpc$. \citet{Cammelli2025} used the constant \diso{} models coupled with a semi-analytic galaxy formation and evolution prescription \citep{Fontanot2020} to study the galactic properties associated with the seeded halos. Based on their $z\sim0$ occupation fraction, galaxy stellar mass function, and black hole mass function results, they concluded that models with $d_{\rm iso}<75~\rm kpc$ are favoured. This is consistent with the range that we obtain for our feedback-regulated \diso{}, and in particular with our fiducial result.


\subsection{Merger rates and dual AGN fraction}

\begin{figure}
    \includegraphics[width=\linewidth]{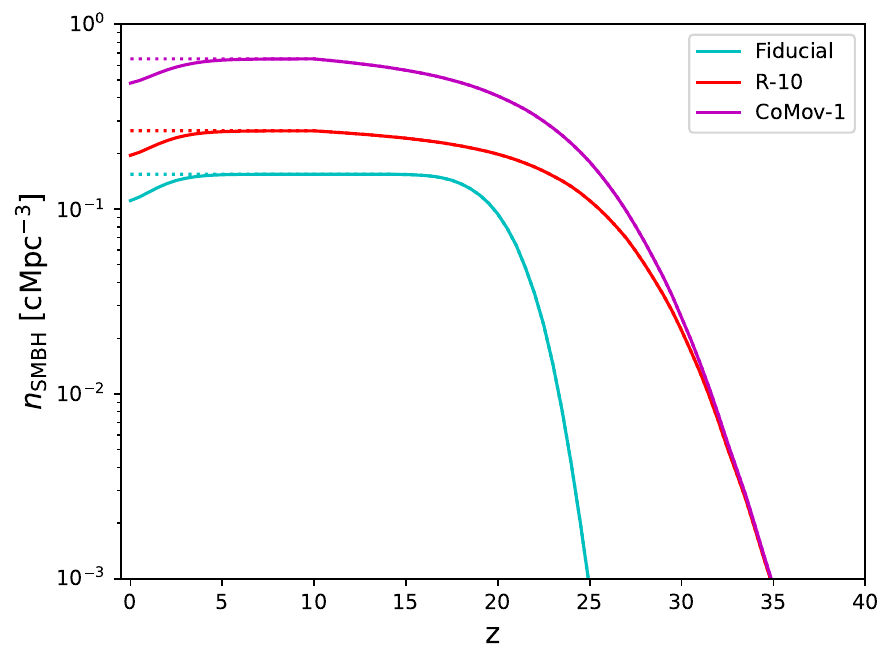}
    \caption{Co-moving number density of SMBH seeds as a function of redshift for the three models evolved to $z=0$. The dotted lines include all seeds present in the simulation box, while the solid lines account for halo mergers.}
    \label{fig:mergers_z0_all}
\end{figure}

\begin{figure}
    \includegraphics[width=\linewidth]{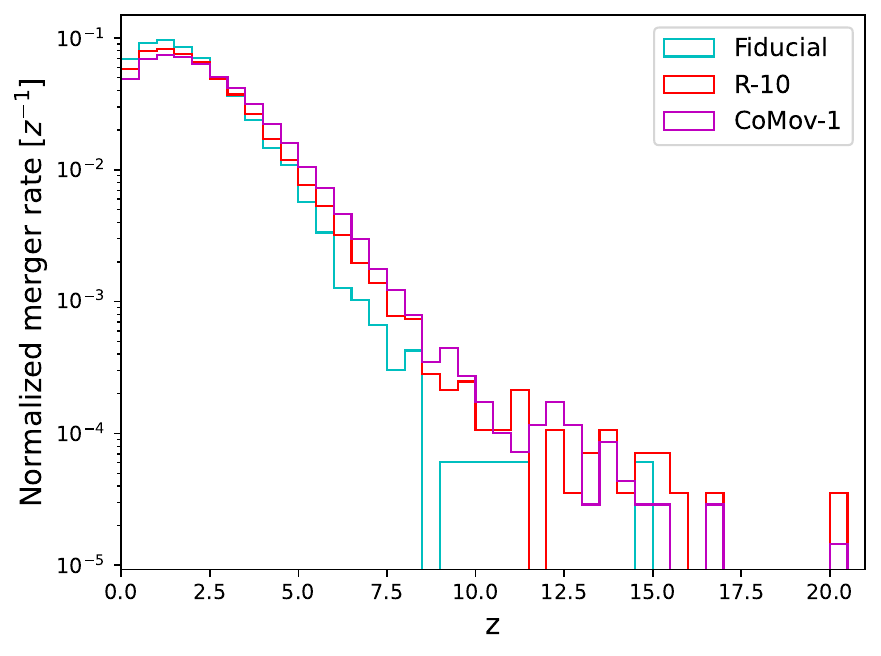}
    \includegraphics[width=\linewidth]{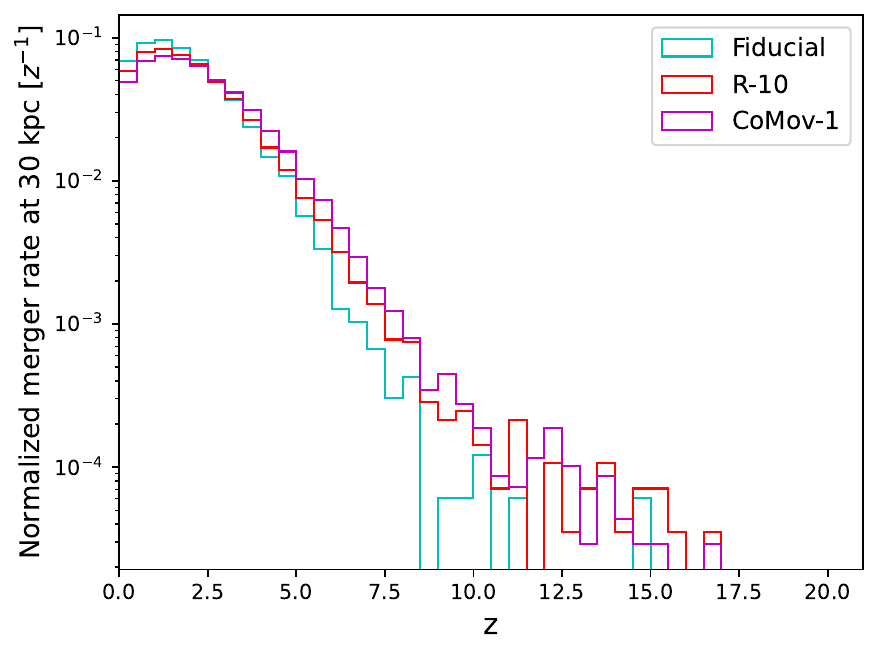}
    \caption{Merger rates as a function of redshift for the three models evolved to $z=0$. \textit{Top:} halo merger events occurring at the separation distances shown in Fig.~\ref{fig:mergers_dist_all}. \textit{Bottom:} same as the top panel, but each merger event was delayed by the free-fall time necessary to reach 30~kpc from the initial halo separation distance.}
    \label{fig:merger_rate}
\end{figure}

Since our models allow us to follow the mergers of dark matter halos down to $z=0$, we consider their implications for occurrences of dual AGN and SMBH merger events. In addition to the fiducial model that was already shown in Figures~\ref{fig:num_dens_fiducial} and~\ref{fig:num_dens_fiducial_data}, we follow the halo merger events of two other models (R-10 and CoMov-1) down to $z=0$. This allows us to make a more robust analysis of the merger events of the seeded halos. The co-moving number densities of the three models are shown in Figure~\ref{fig:mergers_z0_all}. In the top panel of Figure~\ref{fig:merger_rate} we show the halo merger rate, normalized to the total number of seeded halos in each model. We can see that the three models have a very similar distribution of halo mergers as a function of redshift, as well as a similar total fraction of seeded halos that merge ($\sim 25-30\%$).

In order to consider the dual AGN fraction, we first need to define it. As discussed in Section~\ref{sec:mergers}, two halos are considered to merge when their separation becomes smaller than the distance proportional to the Lagrangian radius of the more massive one (see eq.~\ref{eq:merg-dist}). After this separation is reached, our code can no longer track the two halos as individual objects. In reality, the merger event of the galaxies contained within these halos will happen with some time delay, and there will be subsequent delay between the galaxy merger and the SMBH merger. Modelling both of these delays contains a large degree of uncertainty and physical processes that are not included in our simulation. Therefore, here we consider that the dual AGN phase begins when the halos 
reach a separation of 30~kpc \citep[similar to observations studies, e.g.][]{Silverman2020}, and it lasts for a fixed amount of time, $t_{\rm AGN}$. In order to have the uniform AGN separation of 30~kpc, we calculate the time that it takes for the galaxy in the less massive halo to free-fall within the gravitational potential of the more massive one. This time delay (individual for each merger event, but typically around $2.6\times10^5\:$yr) is then used to determine the beginning of the dual AGN phase for each merging pair. This gives us the normalized halo merger rate at a fixed separation of 30~kpc shown in the bottom panel of Fig.~\ref{fig:merger_rate}.

\begin{figure}
    \includegraphics[width=\linewidth]{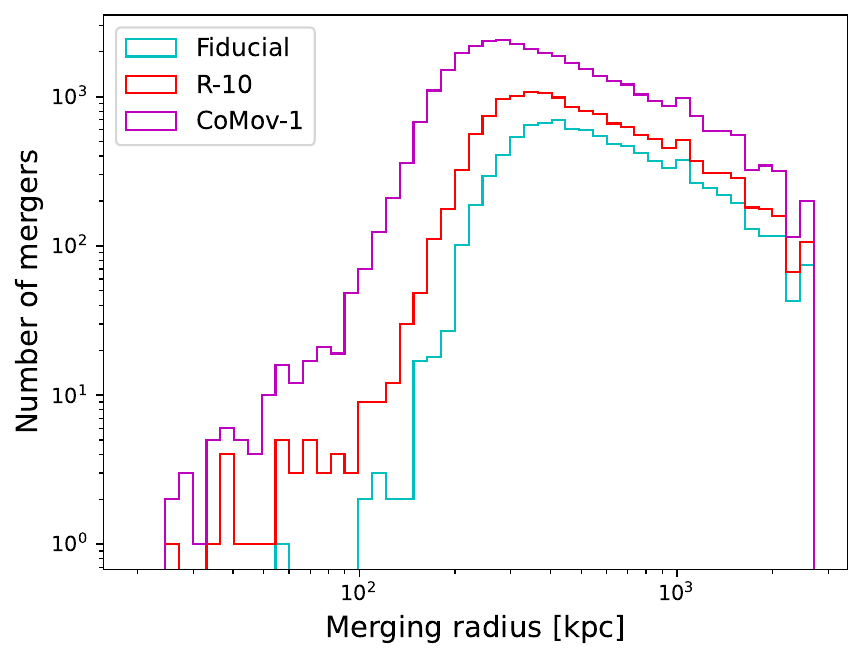}
    \includegraphics[width=\linewidth]{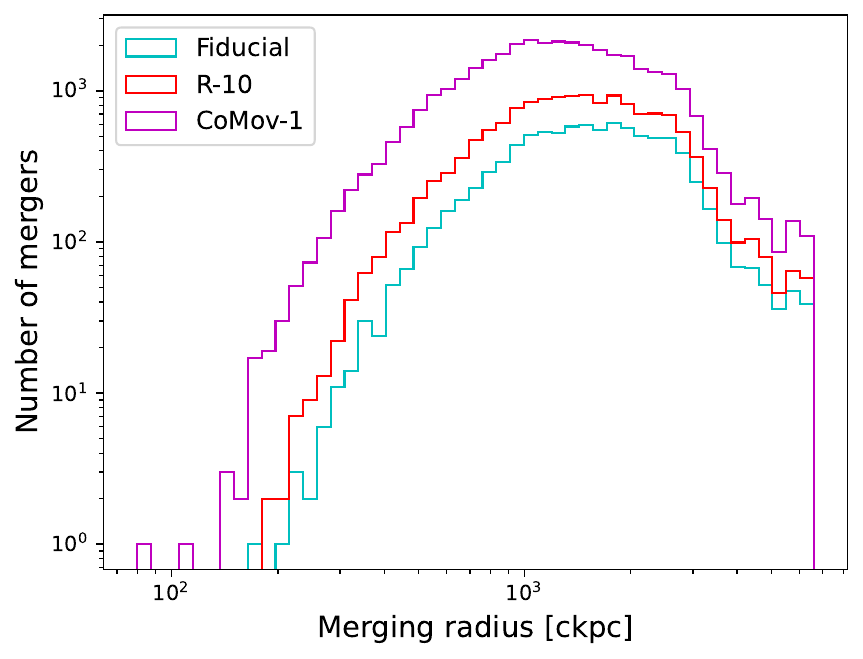}
    \caption{Histograms of halo merger distances for the models evolved to $z=0$. The distances are computed using eq.~\ref{eq:merg-dist} for each merger event.}
    \label{fig:mergers_dist_all}
\end{figure}

Note that the typical separation of the halos (and hence the AGN) at the time of the halo merger is a few $\times 100~\rm kpc$ (see Figure~\ref{fig:mergers_dist_all}). However some of the mergers can begin at a few Mpc. This spread of merger distances may appear concerning when it is larger than the feedback distance in co-moving units (see the bottom panel of Figure~\ref{fig:mergers_dist_all}). For reference, the fiducial model and the R-10 model have \diso{} of $\sim 1.3~\rm cMpc$, and the CoMov-1 model has \diso{} of $1~\rm cMpc$. However, this is mainly a concern for mergers occurring at $z>10$ where we perform the seeding algorithm. The fraction of such merger events for all models is $0.01-0.06\%$ of the total number of mergers for the respective model. Therefore we do not expect such mergers to affect our results significantly. We remind the reader that the seeding is performed on a high resolution simulation, while the mergers are followed on a low resolution simulation that matches the initial conditions of the high resolution one. A low resolution merger occurs at a greater distance than its high resolution counterpart due to the fact that the smallest resolved halo is more massive. Within the high resolution simulation we do not see merger events of seeded halos at $z\gtrsim10$. 

\begin{figure}
    \includegraphics[width=\linewidth]{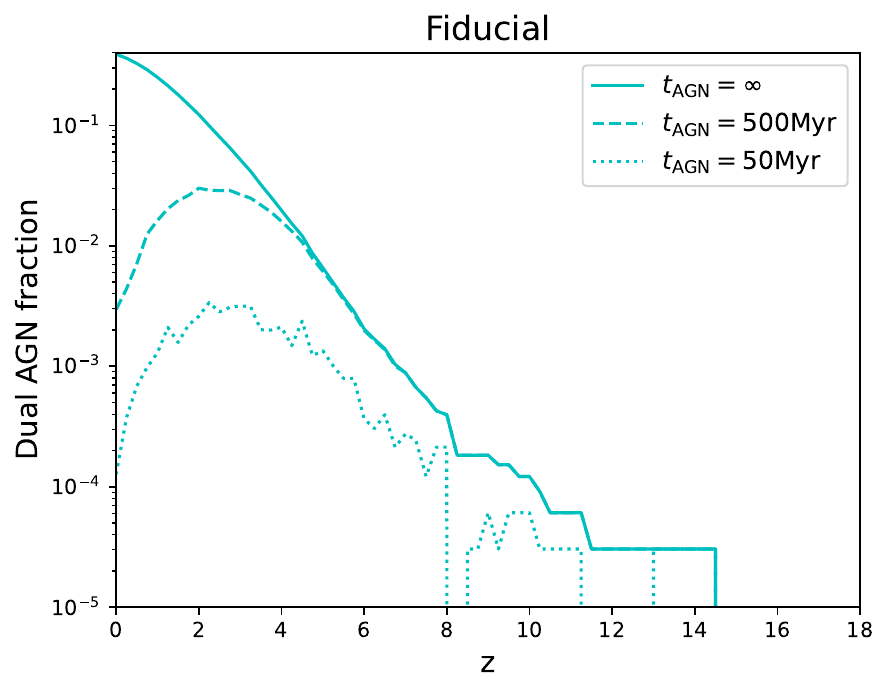}
    \includegraphics[width=\linewidth]{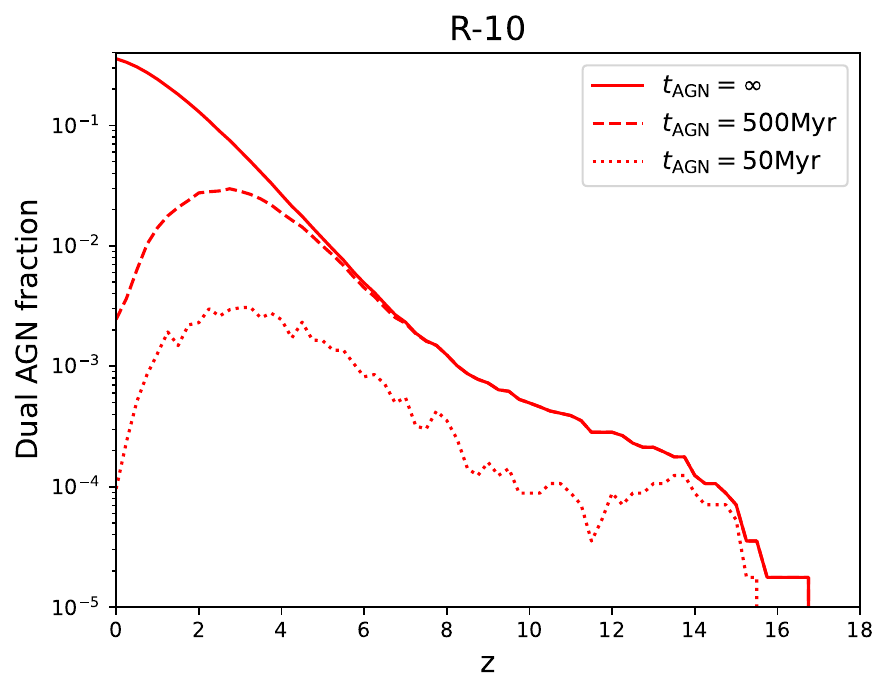}
    \includegraphics[width=\linewidth]{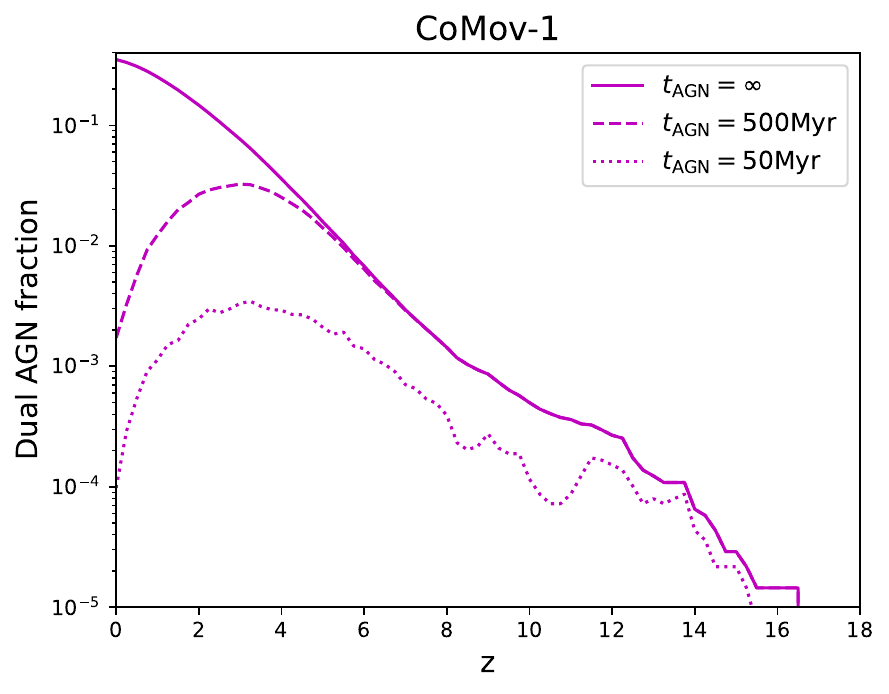}
    \caption{Dual AGN fraction as a function of redshift. For each seeding model, we consider a fixed duration of the AGN phase ($t_{\rm AGN}$).}
    \label{fig:dual_agn_all}
\end{figure}

Figure~\ref{fig:dual_agn_all} shows the dual AGN fraction for each of the models assuming the limit of infinitely long duration of the dual AGN phase, as well as $t_{\rm AGN} = 500~\rm Myr$ and $t_{\rm AGN} = 50~\rm Myr$. Once again, we see good agreement between the different seeding models. In the case of infinitely long dual AGN phase, the dual AGN fraction becomes the cumulative distribution of the halo merger histograms in bottom panel of Figure~\ref{fig:merger_rate}. As $t_{\rm AGN}$ is lowered, the peak of the dual AGN fraction becomes lower ($\sim 0.3-3\%$) and it shifts to $z\sim 2-3$. A key prediction of our model is therefore the very low occurrence of dual AGN at high redshift.

\citet{Silverman2020} found a dual fraction of $0.26 \pm 0.18\%$ for quasars up to $z\sim4.5$. Their results show a flat redshift distribution with a likely downturn at $z\sim4$. Recent observational results by \citet{Jiang2026} found a lower fraction ($0.056 \pm 0.03\%$), but also with a flat distribution between $z=0.5-4.5$. While our dual AGN fraction tends to decrease at $z<2$, we find good agreement between the observed values and our models with $t_{\rm AGN}=50~\rm Myr$. 

\begin{figure}
    \includegraphics[width=\linewidth]{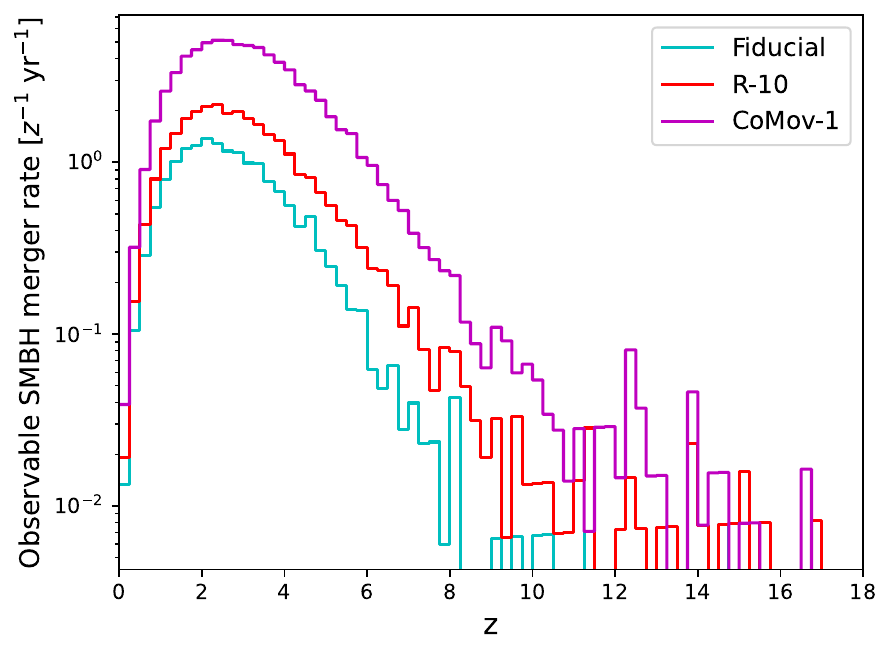}
    \caption{Observable SMBH merger rate as a function of redshift in the limit of fast mergers.}
    \label{fig:obs_merg_rate}
\end{figure}

In addition to the dual AGN fraction, we can compute the predicted rate of SMBH mergers. Since our model considers heavy ($\sim 10^5~\rm M_{\odot}$) SMBH seeds, their mergers will typically be detectable by the planned Laser Interferometer Space Antenna (LISA), which will be sensitive to mergers in the range of $\sim 10^4-10^8\:M_{\odot}$ out to $z\sim10$ \citep{Amaro-Seoane2023}. The modelling of the time delay between a halo merger and a subsequent merger of an SMBH binary involves many complex physical considerations \citep[e.g.,][]{Milosavljevic2001,Sesana2006,Tanaka2009,Langen2025,Singh2026} and is beyond the scope of this work. Therefore, here we only consider an upper estimate of the merger rate in the limit of instantaneous SMBH merging at the beginning of the dual AGN phase. 
We divide the redshift range where mergers occur in bins (up to $z=18$), and for each bin we calculate the number of merging pairs based on the normalised merger rate shown at the bottom panel of Figure~\ref{fig:merger_rate}, the co-moving number density (Figure~\ref{fig:mergers_z0_all}) and the volume of the Universe in the redshift bin relative to the volume of the simulation box. We divide the number of merging pairs by the time-dilated duration of the redshift bin to get a merger rate contribution from the bin (see Figure~\ref{fig:obs_merg_rate}) and add these contributions to obtain the predicted merger rate. 
This gives us merger rates in the range of a few per year (Fiducial: $\sim4~\rm yr^{-1}$, R-10: $\sim7~\rm yr^{-1}$, CoMov-1: $\sim19~\rm yr^{-1}$).


\section{Conclusions}
\label{sec:conclusions}

We have developed a feedback-regulated seeding model of supermassive Pop III.1 stars in the early universe, and applied it to a cosmological dark matter simulation. Within our model, the Pop III.1 stars form isolated from each other, with separations corresponding to the sizes of their relic HII regions. The sizes of the HII regions may depend on redshift as well as stellar parameters, such as the hydrogen ionising photon rate, the lifetime of the stars and their formation delays. These Pop III.1 stars are considered to be the direct progenitors of heavy ($10^5\:M_{\odot}$) SMBH seeds. The main conclusions of our study are as follows.


\begin{itemize}
    \item SMBHs produced by this Pop III.1 mechanism are seeded rapidly at high redshifts, with the process being complete by $z\gtrsim16$.
    \item The final co-moving number density of seeds depends on the parameter choices, and it could span a broad range of values ($\sim 0.03-0.7~\rm cMpc^{-3}$), with the fiducial model reaching $\sim 0.16~\rm cMpc^{-3}$.
    \item The models presented here can easily account for recent observations of AGN at at $z\gtrsim7$, as well as estimates of AGN number density from, e.g., the Hubble Ultra Deep Field \citep{Hayes2024,Cammelli2025b}.
    \item We found that allowing for the expansion of R-type HII regions from Pop III.1 stars resulted in a seeding separation distance that is approximately constant in co-moving distance, as opposed to in proper distance as considered in previous work. For reference, our fiducial model has a seeding separation of $\sim 1.3~\rm cMpc$. 
    \item Thus a key prediction of the Pop III.1 model is that there is an early phase of ``flash'' ionization, which has implications for observations of the CMB \citep{Tan2025,TanKomatsu2025}. Our work presents a more detailed treatment of this phase of early ionization. In particular, to the extent that Pop III.1 stellar parameters are relatively uniform, there is a characteristic scale associated with ionized bubbles in the early universe, which has further implications for observations of the CMB and high redshift atomic H.
    \item Our models predict seed distribution with initially low clustering. Therefore, mergers of seeded halos happen predominantly at $z\leq4$. Approximately $25-30\%$ of all seeded halos end up undergoing mergers with each other.
    \item We estimated the dual AGN fraction produced by our models to be in the range $0.1-3\%$ for $z\approx1-5$. The dual AGN fraction decreases at higher redshifts: in the fiducial case it is $\lesssim0.3\%$ for $z>6$.
    \item In the limit of fast SMBH binary mergers, our models predicted upper limit estimates of LISA detection rates of a few tens per year and fiducial rates of several per year. 

\end{itemize}

Some of the main sources of uncertainty in the model arise from the conditions for Pop III.1 minihalo formation and the properties of their resulting stars. By varying the ionizing luminosity and the lifetime of the Pop~III.1 stars, we can alter the final number density by close to an order of magnitude in each direction. However, even then our model predicts a high number density of SMBHs that are already in place by $z\gtrsim20$. These considerations motivate the need for more detailed studies of the formation and evolution of Pop III.1 supermassive stars powered by dark matter annihilation \citep[e.g.,][]{Rindler-Daller2015,Nandal2025,Topalakis2025}.

\section*{Acknowledgements}

M.A.P. acknowledges funding from a Chalmers Initiative on Cosmic Origins (CICO) postdoctoral fellowship. J.C.T. acknowledges funding from CICO, the Virginia Initiative on Cosmic Origins (VICO), and the Virginia Institute for Theoretical Astrophysics (VITA), supported by the College and Graduate School of Arts and Sciences at the University of Virginia. J.S. acknowledges a financial contribution from the Bando Ricerca Fondamentale INAF 2022 Large Grant, ‘Dual and binary supermassive black holes in the multi-messenger era: from galaxy mergers to gravitational waves’ and from the INAF Bando Ricerca Fondamentale INAF 2024 Large Grant, ‘The Quest for dual and binary massive black holes in the gravitational wave era’.

\section*{Data Availability}

The data underlying this article will be shared on reasonable request to the corresponding author.



\bibliographystyle{mnras}
\bibliography{phys-diso} 




\appendix



\bsp	
\label{lastpage}
\end{document}